\documentclass[aps,prl,reprint,preprintnumbers,superscriptaddress,showpacs]{revtex4-1}

\usepackage{hyperref}
\usepackage{amsmath,amssymb,amsfonts}
\usepackage[T1]{fontenc} 
\usepackage{graphicx}
\usepackage{physics}

\usepackage{color}

\newcommand{\Z}{\mathbb{Z}}
\newcommand{\R}{\mathbb{R}}
\def\C{\mathbb{C}}
\renewcommand{\>}{\rangle}
\newcommand{\<}{\langle}
\newcommand{\up}{\uparrow}
\newcommand{\down}{\downarrow}
\newcommand{\sgn}{\mathop{\text{sgn}}}
\newcommand{\Laplace}{\Delta}
\newcommand{\del}{{\partial}}

\newcommand{\numberthis}{\addtocounter{equation}{1}\tag{\theequation}}

\newcommand{\x}{{\mathbf x}}
\newcommand{\A}{{\mathbf A}}
\newcommand{\E}{{\mathbf E}}
\newcommand{\B}{{\mathbf B}}
\newcommand{\Chiral}{{\hat C}}
\renewcommand{\H}{{\hat H}}
\newcommand{\T}{{\hat T}}
\newcommand{\rr}{{\hat\rho}}
\newcommand{\psia}{{\psi_\up}}
\newcommand{\eps}{\varepsilon}

\begin{document}

\title{Time-reversal-symmetric topological magnetoelectric effect
\\ in three-dimensional topological insulators}

\author{Heinrich-Gregor Zirnstein}
\author{Bernd Rosenow}
\affiliation{Institut f\"ur Theoretische Physik, Universit\"at Leipzig, D-04103 Leipzig, Germany}

\date{\today}

\begin{abstract}
One of the hallmarks of time-reversal-symmetric topological insulators in three dimensions is the topological magnetoelectric effect (TME). So far, a time-reversal breaking variant of this effect has attracted much attention, in the sense that the induced electric charge changes sign when the direction of an externally applied magnetic field is reversed. Theoretically, this effect is described by the so-called axion term. Here, we discuss a time-reversal-symmetric TME, where the electric charge depends only on the magnitude of the magnetic field but is independent of its sign. We obtain this nonperturbative result both analytically and numerically, and suggest a mesoscopic setup to demonstrate it experimentally.
\end{abstract}

\maketitle

\noindent
\emph{Introduction}.
Time-reversal-symmetric (TRS) topological insulators (TIs)~\cite{Hasan:2010ku,Qi:2011hb} are a fascinating class of electronic materials with  insulating bulk and topologically protected surface states, which are either gapless, break a symmetry, or feature topological order~\cite{Vishwanath:2013fv}.
Evidence for the existence of such surface states comes from spin textures observed in photoemission experiments~\cite{Xia:2009fn,Zhang:2009ks} and from the observation of a half-integer quantum Hall effect%
~\cite{
Brune:2011hi,Chang:2013dd,Xu:2014eh,Yoshimi:2015ep}%
.

From a theoretical point of view, the  hallmark response of TRS TIs in three dimensions (3D) is the topological magnetoelectric effect~\cite{Qi:2008eu, Essin:2010gy}. So far, a time-reversal (TR) breaking variant of this effect has attracted much attention. When TR is broken by, say, a magnetic  coating with Zeeman coupling to the TI surface, a Hall conductivity $\sigma_{xy}=\frac{\tilde{\theta}}{2\pi} \frac{e^2}{2\pi}$ arises, where $\tilde{\theta}$ is quantized to $\tilde{\theta}=\pm \pi$.  
Then, the insertion of a magnetic flux tube gives rise to the accumulation of a charge $|Q|=e/2$ per flux quantum $\Phi_0=h/e$.
Importantly, $\sgn(Q)$ depends on the direction of the magnetic field inside the flux tube, i.e.,~the response is not TRS. 
A consequence of the surface Hall conductivity is quantized Kerr and Faraday rotations~\cite{Tse:2010bd,Maciejko:2010dx}, which have recently been confirmed experimentally~\cite{Wu:2016em,Okada:2016ck,Dziom:2016uh}.

In the presence of TRS, the linear magnetoelectric response vanishes \cite{Landau:1984ui,Note17}.
Thus, strictly speaking, all variants of the topological magnetoelectric effect are nonlinear effects, as they require an additional perturbation, say, a Zeeman coupling on the surface as above.
The absence of a linear magnetoelectric response may seem to be at odds with the fact that the bulk of a 3D TRS TI has been characterized by the so-called axion action~\cite{Qi:2008eu, Essin:2010gy} $S_{{\theta}}=\frac{{\theta}}{2\pi} \frac{e^2}{2\pi} \int d^3x\,dt\, \E\cdot\B$.
Under a TR transformation, $\E \to \E$, $\B \to - \B$, and $S_{{\theta}} \to - S_{{\theta}}$.
Classically, this action breaks TRS, but quantum mechanically, only the Feynman amplitude $e^{i S_{{\theta}}}$ needs to be symmetric.
If the electronic wave functions and electromagnetic fields satisfy periodic boundary conditions, one can show that this integral is quantized to integer multiples of $4 \pi^2/e^2$~\cite{Milnor:1974wr,Vazifeh:2010iu}. This implies that $S_{{\theta}}= {\theta}$ modulo $2 \pi$, hence $\theta=\pm \pi$ would respect TRS also.
Now, for a TI with boundaries, by using partial integration, $S_{{\pm\pi}}$ can be converted into the surface quantum Hall term discussed above, and it seems naively that the axion action indeed describes a magnetoelectric effect.
However, this is not the case, as the surface response due to $S_{\pm\pi}$ is canceled by additional contributions from the surface states, in this case the parity anomaly, to restore TRS~\cite{Mulligan:2013cz, Zirnstein:2013wk,Witten:2016fu}. 

%

\begin{figure}[t]
\begin{center}
\includegraphics[width=0.45\textwidth]{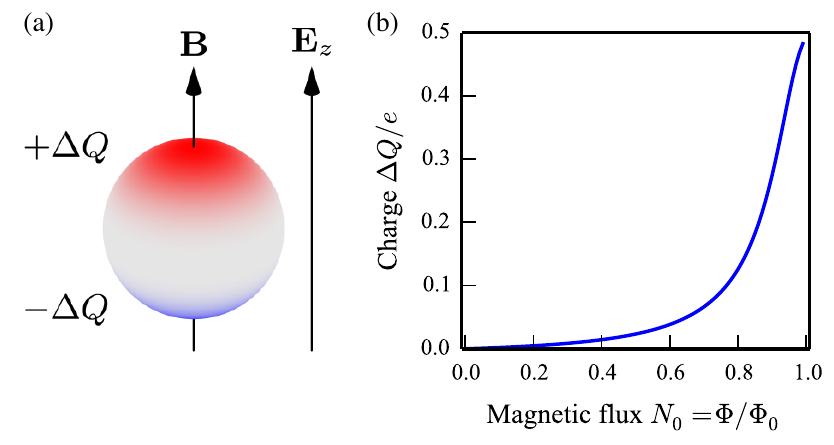}
\end{center}
\caption{%
Spherical topological insulator threaded by a thin magnetic flux tube with flux $\Phi$ and subject to an electric field $\E_z$ in vertical direction.  Insertion of one flux quantum induces a charge $\Delta Q = e/2$. 
(a) Geometry.
(b) Analytical result for the charge difference in the top hemisphere, $\Delta Q(\mathbf{E}_z,\Phi)$, in the thin flux tube limit for an external electric field $\mathbf{E}_z$, giving rise to a potential energy difference $e2RE_z$ between top and bottom, with $e2RE_z = 0.2 v_F/R$. Here, $R$ denotes the radius of the sphere and $v_F$ is the Fermi velocity on the surface.
\label{figure-setup-sphere}}
\end{figure}

In this work, we describe a TR-symmetric topological magnetoelectric effect.
This means that the accumulated electric charge depends only on the magnitude of the magnetic flux but is independent of its sign (magnetic field direction).
In particular, we consider a spherical TI threaded by a thin magnetic flux tube, and subject to a small uniform electric field [see Fig.~\ref{figure-setup-sphere}(a)].
If $Q_{\text{top}}(\mathbf{E}_z,\Phi)$ denotes the total charge on the top half, we show that the additional charge due to the insertion of one flux quantum is
$\Delta Q(\mathbf{E}_z,\pm\Phi_0) \equiv Q_{\text{top}}(\mathbf{E}_z,\pm\Phi_0) - Q_{\text{top}}(\mathbf{E}_z,0) = +(e/2) \sgn \mathbf{E}_z$.
This response is TRS, in contrast to the TR-breaking surface quantum Hall effect%
~\cite{
Brune:2011hi,Chang:2013dd,Xu:2014eh,Yoshimi:2015ep,
Nomura:2011ke,Konig:2014gb,
Lee:2009do,Vafek:2011ie,
Nogueira:2016em}
described above.
In addition, we numerically analyze a lattice model, and suggest a mesoscopic setup to demonstrate this effect experimentally. These results are not related to the wormhole effect~\cite{Rosenberg:2010dj}, which only occurs when the diameter of the flux tube is much smaller than the lattice spacing. Here, we instead consider a flux tube that covers many plaquettes, so that the surface states of the TI can be described by a continuum two-dimensional (2D) Dirac Hamiltonian.
Then, on an infinite planar surface, inserting one flux quantum $\Phi_0$ will give rise to a single, spin-polarized zero-energy state localized (power law) at the tube~\cite{Aharonov:1979,Jackiw:1981cq,Polychronakos:1987}.
Our treatment of a finite geometry goes significantly beyond this result.

\emph{Physical picture}.
The fractional charge $e/2$ arises because an occupied delocalized eigenstate is transformed into an occupied localized eigenstate when adiabatically inserting one flux quantum:
In a spherical geometry, the delocalized state has equal weight in both hemispheres, whereas the localized state is contained in one of them; this corresponds to a change of charge in that hemisphere by $e/2$.
The presence of a small background field $\mathbf{E}_z$ is important for this to happen: Flux insertion generates two localized states, one at each pole, which would otherwise hybridize with each other, resulting in delocalized eigenstates again, and no charge difference would be observable. Only when the energy difference between the localized states, due to $\mathbf{E}_z$, is larger than the hybridization, one localized state will be occupied, and the other one empty.

\begin{figure}
\begin{center}
\includegraphics{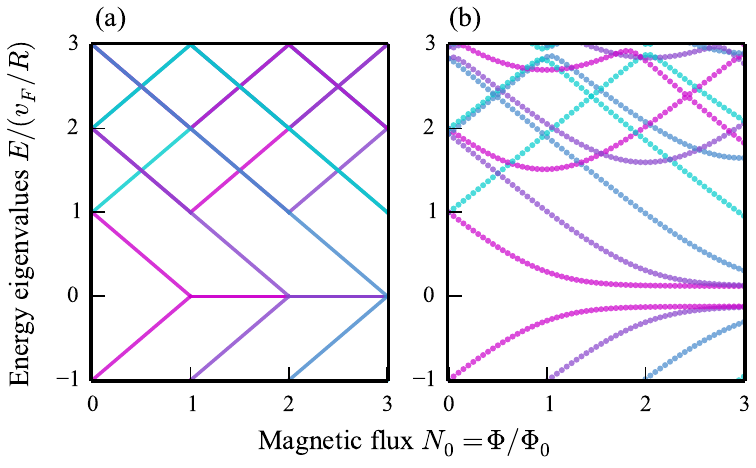}
\end{center}
\caption{%
Low-energy spectrum of a TI with a magnetic flux tube.
(a) Analytical results for a spherical geometry in the limit of a thin flux tube. Angular momentum $m=\pm\frac12,\pm\frac32,\dots$ is encoded by different colors.
(b) Numerical results for the lattice model, Eq.~\eqref{eq-hamiltonian-lattice}, in a cuboid geometry with $10\times 10\times 8$ sites, a flux tube spanning $3\times 3$ plaquettes, and parameters $\lambda=1$, $t=2$, and $\kappa=4t$.
The hybridization of the low-energy states after one flux quantum is due to the finite size of the flux tube.
\label{figure-spectrum}%
}\end{figure}

\begin{figure}
\begin{center}
\includegraphics{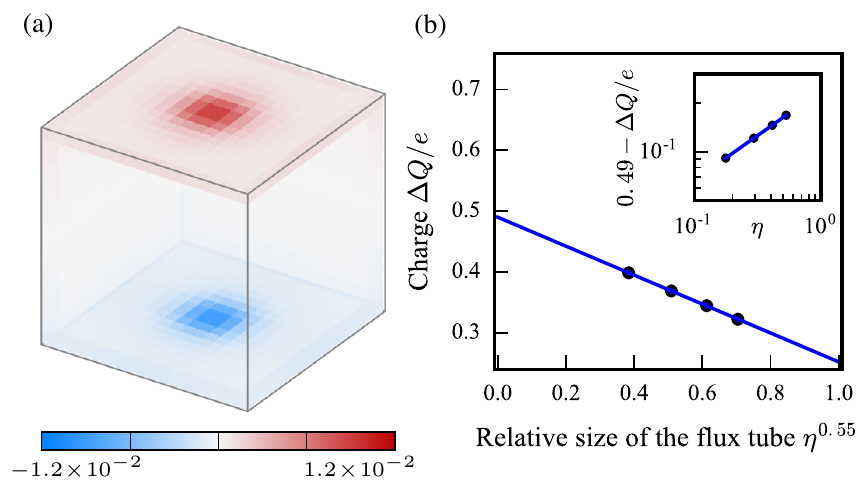}
\end{center}
\caption{%
Charge induced in the TI lattice model, Eq.~\eqref{eq-hamiltonian-lattice}, in a cuboid geometry with $18\times 18\times 18$ sites in the presence of a magnetic flux tube of unit flux $\Phi/\Phi_0=1$.
Parameters are $\lambda=2.43$, $t=\lambda$, $\kappa=4t$, and $eU=0.3\lambda$.
(a) Charge distribution for a flux tube spanning $5\times 5$ plaquettes.
(b) Scaling of the total induced charge in the top half of the cube depending on the relative size $\eta$ of the flux tube ($\eta \propto \text{tube diameter}$). The inset shows that the charge scales according to a power law.
\label{figure-charge-numerical}%
}\end{figure}

In order to evaluate the induced charge, we compute the two eigenstates with energies closest to the Fermi level for each value of the flux $\Phi$ between zero and $\Phi_0$, and then evaluate the effect of an external electric field $\mathbf{E}_z$ in the subspace spanned by these states. This projection to a low-energy subspace becomes exact if one first takes the limit of  an infinitesimally thin flux tube [see Fig.~\ref{figure-spectrum}(a)], and then considers an infinitesimal electric field $\mathbf{E}_z$. In this way, the infinitesimal $\mathbf{E}_z$ does not polarize the system for zero flux, but due to the order of limits described above, at $\Phi=\Phi_0$ the level splitting caused by $\mathbf{E}_z$  is much larger that the hybridization energy. In a numerical calculation, both the flux tube diameter and the electric field are small but finite, and the exact analytic result is recovered by finite-size scaling [see Fig.~\ref{figure-charge-numerical}(b)].

The two lowest-energy eigenstates are superpositions of spin-polarized wave functions $\eta_{\text{n}}(\Phi,\x)$ and $\eta_{\text{s}}(\Phi,\x)$ located at the north and south pole; explicit expressions are derived later.
Projecting to these states, the Hamiltonian becomes $\H = \Delta s_x + V s_z$ where $s_x$, $s_z$ are Pauli matrices, $\Delta$ is the hybridization energy, and $V$ is the projected potential energy $V(\Phi) = e d(\Phi) \mathbf{E}_z$ with $d(\Phi) = \<\eta_{\text{n}}|\x_3|\eta_{\text{n}}\>$ the dipole moment.
The charge response is $\Delta Q(\mathbf{E}_z,\Phi) = \rho(\mathbf{E}_z,\Phi) - \rho(\mathbf{E}_z,0)$ with
\begin{equation}
    \label{eq-two-level-charge-density}
    \rho(\mathbf{E}_z,\Phi)
    = \frac{e}2 \frac{w(\Phi)}{\sqrt{V(\Phi)^2 + \Delta(\Phi)^2}} V(\Phi)
,\end{equation}
and $w(\Phi) = \int_{\text{top}} d\x\,[\abs{\eta_{\text{n}}(\Phi,\x)}^2 - \abs{\eta_{\text{s}}(\Phi,\x)}^2]$.
In the limit of a thin flux tube, the hybridization equals the energy $\Delta(\Phi) = (\Phi/\Phi_0-1)v_F/R$ for $\Phi<\Phi_0$
where $v_F$ is the Fermi velocity and $R$ the radius of the sphere.
The corresponding charge response is depicted in Fig.~\ref{figure-setup-sphere}(b).
Since we have $\Delta(\Phi) \to 0$ and $w(\Phi)\to 1$ for $\Phi\to\Phi_0$, an infinitesimally small electric field is sufficient to lift the degeneracy, and we find that $\Delta Q (\mathbf{E}_z,\Phi_0) = +(e/2)\sgn \mathbf{E}_z$ for $\mathbf{E}_z\to 0$.

\emph{Fractional charge}.
The appearance of a half-integer charge due to localized zero-energy states can be interpreted as an instance of charge fractionalization~\cite{Jackiw:1981cq,Krive:1987,Leinaas:2009vx,Leinaas:2009ks}.
To make the connection to condensed matter, we consider a general lattice Hamiltonian $\H(A_0,\A)$ which depends on an electric potential $A_0$ and a vector potential $\A$, and has $n$ internal degrees of freedom per site $\x$.
At zero temperature and chemical potential, the expectation value of the charge density is
\begin{equation}
    \label{eq-charge-density}
    \<\rr(\x)\>_{A_0,\A} = (-e)\sum_{E_\alpha \leq 0} |\eta_\alpha(\x)|^2
,\end{equation}
where $E_\alpha$ are the energy eigenvalues of $\H(A_0,\A)$, and $\eta_\alpha(\x)$  the corresponding eigenfunctions. 
We now focus on Hamiltonians with a chiral pseudosymmetry, described by a local operator $\Chiral$ anticommuting with the Hamiltonian at zero electric field, $\{\H(A_0=0,\A), \Chiral\}= 0$.
Microscopically, such a pseudosymmetry can be realized for Hamiltonians defined on a bipartite lattice, with hoppings only between sublattices.
Anticommutation implies that eigenstates come in pairs with positive and negative energy $E=E_\alpha=-E_\beta$, and locality means that the corresponding eigenfunctions have equal probabilities, $|\eta_\beta(\x)|^2 = |\Chiral\eta_\alpha(\x)|^2 = |\eta_\alpha(\x)|^2$. In particular, each probability can be expressed as half the sum of a positive and a negative energy eigenstate, $|\eta_{\alpha}(\x)|^2 = \frac12(|\eta_{\alpha}(\x)|^2+|\eta_{\beta}(\x)|^2)$.
In the absence of zero-energy states, whose consequences will be discussed shortly, $E_\alpha \neq 0$, we can apply the above decomposition to Eq.~\eqref{eq-charge-density}, and find that the charge density can be expressed as $1/2$ times the sum over all eigenstates. But this sum corresponds to all bands completely filled, hence the electron density in the absence of an electric field is \emph{spatially uniform} and has the constant value $\<\rr(\x)\>_{A_0=0,\A}=(-e)n/2$.

The charge deviation from a reference configuration without fields is $\delta\<\rr(\x)\>_{A_0,\A} = \<\rr(\x)\>_{A_0,\A} - \<\rr(\x)\>_{A_0=0,\A=0}$.
Being constant, the reference charge density is $1/2$ times the sum over any complete set of states, giving
\begin{equation}
    \label{eq-eta-invariant}
    \delta\<\hat\rho(\x)\>_{A_0,\A}
    =
    (-e)\left[
    \frac12 \sum_{E_\alpha \leq 0} |\eta_\alpha(\x)|^2
    - \frac12 \sum_{E_\beta > 0} |\eta_\beta(\x)|^2
    \right]
.\end{equation}
In high-energy physics,
this expression defines the vacuum charge~\cite{Jackiw:1981cq, Krive:1987}.
Here, an occupied state contributes a total charge of $\delta Q=-e/2$, an empty state contributes $\delta Q=+e/2$, while zero-energy states are ambiguous and can be attributed to either sum.
In our setup, the magnetic flux gives localized zero-energy states, and this formula now describes a physical, localized charge deviation $\delta Q=\mp e/2$ whose sign is determined by the occupation of the states selected by the infinitesimal electric field $\mathbf{E}_z$.

\emph{Analytical results}.
Since the low-energy states of a TI are localized on the surface
\setcounter{footnote}{16}\footnote{See Supplemental Material at \dots{} for details regarding time-reversal symmetry and the calculation of the eigenstates in the presence of a thin magnetic flux tube.},
we consider the Dirac Hamiltonian on a sphere with radius $R$, \cite{Imura:2012cz,Lee:2009do,Parente:2011fj}
\begin{equation}
    \label{eq-hamiltonian}
    \H = \begin{pmatrix}
        0 & h^+ \\
        h^- & 0
    \end{pmatrix}
,\
h^{\pm} = \mp \left(\del_\theta + \frac12 \cot \theta\right) + \frac{i\del_\phi}{\sin \theta} + eR\A_\phi
,\end{equation}
where $\phi\in[0,2\pi)$, $\theta\in[0,\pi]$ are  spherical coordinates, and energy is measured in units of $v_F/R$ with $v_F$ denoting the velocity of Dirac electrons. We have specialized to a vector potential $\A$ which has only an azimuthal component $\A_\phi(\theta)$, reflecting rotational symmetry around the flux tube. 
We can decompose $\psi(\phi,\theta) = \tilde{\psi}(\theta) e^{im\phi}/\sqrt{R}$ with half-integer angular momentum $m=\pm\frac12,\pm\frac32,\dots$ for  spin-$1/2$ electrons.  In the absence of an external field, $\mathbf{A_\phi}=0$, the energy eigenvalues of the spherical Dirac operator are known to be nonzero integers $E=\pm 1,\pm 2,\dots$ whose multiplicities increase with angular momentum~\cite{Newman:1966ex,AbrikosovJr:2002tx}. It is tempting to incorporate the flux tube by using the Aharonov-Bohm effect and shifting the angular momentum, $m \to m - N_0$~\cite{Lee:2009do}, where $N_0=\Phi/\Phi_0$ is the number of flux quanta, but this approach does not allow the implementation of the correct boundary condition that the wave function $\tilde{\psi}(\theta)$ stays finite as $\theta \to 0,\pi$ for spatial coordinates \textit{within the region of the flux tube}.
\footnote{Ref.~\cite{Lee:2009do} does not mention boundary conditions and concludes that even in the absence of an external magnetic field, the spherical Dirac operator has zero-energy states, $E=0$, which does not agree with our results.}

To model the flux tube, we express the vector potential as
$N_\phi(\theta) = e R \A_\phi(\theta)\sin\theta$
and substitute $x=\cos\theta$.
Then, we choose $N_\phi(x) = N_0 \min\{1,(1-|x|)/\delta\}$, which is equal to the total flux $N_0$ in most of the sphere, but vanishes at the poles. 
For a thin flux tube, we have $0< \delta \ll 1$.
Using that $\H^2 = \text{diag}(h^+h^-,h^-h^+)$,  we only need to solve the eigenvalue equation $h^+h^- \psia = E^2\psia$. Then, the eigenvectors of the original Hamiltonian are obtained as $\psi_{\pm E} = (\pm E\psia, h^-\psia)^T$ if $E\neq 0$. Zero modes, $E=0$, are obtained from $h^-\psi_\up=0$, or $h^+\psi_\down=0$. One finds that 
\begin{align}
\nonumber
    h^+h^- &=
    -\frac{d}{dx}\left[(1-x^2) \frac{d}{dx} \right]
\\ & \quad
    + \frac1{1-x^2} \left[-m + \frac12 x + N_\phi\right]^2
    + \frac{dN_\phi}{dx}
    + \frac12
    \label{eq-hamiltonian-squared}
,\end{align}
resembling the Legendre differential operator.
This is a special case of the Schr\"odinger--Lichnerowicz formula~\cite{Lichnerowicz:1963vc}.
We now use a piecewise ansatz \cite{AbrikosovJr:2002tx,Imura:2012cz}, here shown for angular momentum $m\geq 1/2$:
\begin{equation}
    \label{eq-wave-function-ansatz-main}
    \psia(x) = \begin{cases}
        (1-x)^{\frac12(m-\frac12)} (1+x)^{\frac12(m+\frac12) - \frac{N_0}{\delta}} g_0(x)
            \\
        (1-x)^{\frac12(m-N_0-\frac12)} (1+x)^{\frac12(m-N_0+\frac12)} g(x)
            \\
        (1-x)^{\frac12(m-\frac12) - \frac{N_0}{\delta}} (1+x)^{\frac12(m+\frac12)} g_1(x)
,\end{cases}
\end{equation}
where the pieces are defined for $1-\delta \leq x\leq 1$, $|x| < 1-\delta$, and $-1 \leq x \leq -1+\delta$, respectively.
We emphasize  that the wave function stays finite near the poles. In coordinates $\xi=(1-x)/2$, the eigenvalue equation is equivalent to a set of hypergeometric equations for the functions $g_0(\xi),g(\xi),g_1(\xi)$,  solved by hypergeometric functions $F(a,b,c;\xi)$~\cite{NIST:DLMF}. Abbreviating $c = (m+1/2-N_0)$ and assuming $c\not\in\Z$, the general solution for the middle part is
\begin{align}
    g&(\xi) =  \alpha F\left(c - E, c + E, c, \xi \right)
\nonumber \\ &
        + \beta \xi^{1-c} F\left(1 - E, 1+E, 2-c, \xi \right)
    \label{eq-wave-hypergeometric}
\end{align}
with parameters $\alpha$, $\beta$ that have to be determined by two jump conditions for derivatives, the first one being
\begin{equation}
    g'(\eps)/g(\eps) - g_0'(\eps)/g_0(\eps) = N_0/\eps
\end{equation}
with $\varepsilon=\delta/2$, and the other similar~\cite{Note17}.
To make progress, we now take the limit of a thin flux tube, $\varepsilon\to 0$.
Expanding the solutions $g_0,g_1$ to leading order in $\varepsilon$ while ignoring powers of order $\eps^{|c|}$ or higher, we find~\cite{Note17} that the jump conditions can only be satisfied for energies $E$ that fulfill
\begin{equation}
    \label{eq-surface-energy-levels}
    E = \pm \begin{cases}
        c+n &\text{ if } c>0\\
        0 \text{ or } 1-c+n &\text{ if } c<0
    ,\end{cases}
\end{equation}
where $n=0,1,2,\dots$. This spectrum is visualized in Fig.~\ref{figure-spectrum}(a).

We can now give explicit wave functions for the two eigenstates with energy closest to zero, used in Eq.~\eqref{eq-two-level-charge-density}.
For flux $0 < \Phi < \Phi_0$, they have angular momentum $m=1/2$ and are superpositions $\eta_{\mp} = (\eta_{\text{n}} \pm \eta_{\text{s}})/\sqrt{2}$ of states $\eta_{\text{n}}(\x) = [\psia(x),0]^T$ and $\eta_{\text{s}}(\x) = [0,\psia(-x)]^T$
with
\begin{equation}
    \label{eq-wave-function-two-level}
    \psia(x) = \frac1{\sqrt{\mathcal{N}}} (1-x)^{\frac12 (-\Phi/\Phi_0)}(1+x)^{\frac12 (1-\Phi/\Phi_0)}
\end{equation}
in the region $|x|<1-\delta$ and with normalization $\mathcal{N} = R [\sqrt{\pi}\Gamma(c)/\Gamma(c+1/2) + O(\delta^c)]$.

\begin{figure}[!tp]
\begin{center}
\includegraphics[width=0.3\textwidth]{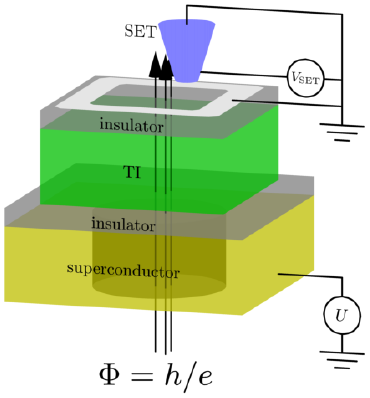}
\end{center}
\caption{%
Illustration of an experimental setup to measure the TRS surface charge. A superconductor focuses the magnetic flux through the TI, while a scanning single-electron transistor (SET) allows sensitive charge measurements~\cite{Venkatachalam:2010kg}. The top and bottom sides of the TI have a voltage difference $U$.
\label{figure-experiment}%
}\end{figure}

\emph{Numerical results}.
To confirm our analytic result, 
we have performed a numerical calculation using a minimal lattice model for a topological insulator~\cite{Qi:2008eu}. This model concerns a four-component fermionic wave function on a cubic lattice. In momentum (Bloch) space, the TR-symmetric Hamiltonian is
\begin{equation}
\label{eq-hamiltonian-lattice}
    \H = -2\lambda \tau_z \sum_{\mu=1}^3 \sigma_\mu \sin(k_\mu)
    + \tau_x \left(\kappa - 2t\sum_{\mu=1}^3 \cos(k_\mu)\right)  ,
\end{equation}
where $\lambda, \kappa, t$ are real parameters, and $\sigma_\mu$ and $\tau_\mu$ are Pauli matrices acting on the spin (respectively, orbital) degrees of freedom. This model is a (strong) topological insulator in the parameter range $2t \leq \kappa \leq 6t$ \cite{Qi:2008eu}. Coupling to the electromagnetic field is achieved by Peierls substitution~\cite{Note17}.

The induced charge in the presence of a flux tube and a small electric field is shown in Fig.~\ref{figure-charge-numerical}. The charge distribution $\delta\<\rr(\x)\>_{A_0,\A}$ is TR symmetric, localized on the surface, and concentrated at the flux tube.
To compare with the analytical results, we need to take the limit of a thin flux tube by a scaling analysis where we fix a system size and shrink the size of the flux tube.
The electric potential has to be smaller than the level spacing, but larger than the hybridization, as the order of limits is important.
We find that the extrapolated value of the charge difference in the top half is $\Delta Q(\Phi_0) = +(0.49\pm0.02)e$, in excellent agreement with our analytical results.
For a topologically trivial insulator (e.g.,\ $6t < \epsilon$, not shown) no significant charge is accumulated at the flux tube.

\emph{Experimental realization}.
A mesoscopic setup for measuring the TR-symmetric charge response is illustrated in Fig.~\ref{figure-experiment}. If we use Bi$_2$Se$_3$ as an example TI \cite{Xia:2009fn} with a Fermi velocity $v_F \sim 5\times 10^5\, \text{m}\,\text{s}^{-1}$ and assume that the system has a diameter of $1 µ\text{m}$, then the level spacing~\eqref{eq-surface-energy-levels} of the surface states, $\Delta E = \hbar v_F/R$, should be on the order of half a $\text{meV}$. This is well below the bulk band gap $E_{\text{gap}}\sim 0.35\, \text{eV}$, but large enough to be comparable to an externally applied voltage.
A thin magnetic flux tube  could be generated  by using a superconductor to focus the magnetic field, or by pinning two magnetic vortices, or a giant vortex \cite{Kanda:2004jh,Cren:2011kj}.
The numerical results, Fig.~\ref{figure-charge-numerical}, indicate that larger tube diameters still yield an appreciable charge response, and the value for a thin tube can be obtained by finite-size scaling. The key signature of a TI is that the charge is a half-integer multiple of the elementary charge, as opposed to an integer multiple for purely 2D materials like graphene. Such an experiment is expected to be challenging but within reach of present day technology. 

\emph{Conclusion}.
While the electromagnetism of TRS TIs in 3D is commonly associated with the axion action, we have argued that this action is inadequate for describing a physical response, in particular one that is TRS. In search of the latter, we have adapted the idea of charge fractionalization from high-energy physics to a condensed-matter setting.
Our main result is that the insertion of a thin flux tube leads to a pair of localized states whose hybridization at one flux quantum is much smaller than the level spacing. Combined with small background electric field, this allows us to adiabatically transform a delocalized eigenstate into a localized eigenstate, giving a TRS charge response of $e/2$ that is, in principle, amenable to experimental detection.

\begin{acknowledgments}
\emph{Acknowledgements}.
We would like to thank J.~Smet and L.~Kimme for helpful discussions. This publication was funded by the German Research Foundation within the Collaborative Research Centre 762 (project B6).
\end{acknowledgments}

\bibliography{bib-fractional}

\onecolumngrid
\clearpage
%
\begin{center}
\textbf{\large Supplemental Material:
Time-reversal-symmetric
\\ topological magnetoelectric effect in three-dimensional topological insulators}
\\[0.4ex] Heinrich-Gregor Zirnstein and Bernd Rosenow
\end{center}
\setcounter{page}{1}
%
\subsection{Time-reversal symmetric response in a finite lattice model}

In this section, we show that any time-reversal symmetric (TRS) finite lattice model has a TRS electromagnetic response. This demonstrates that the linear magnetoelectric effect vanishes for a TRS topological insulator, and that the axion action cannot not describe a TRS response of a finite system.

We consider any finite lattice, with or without boundary, where sites are labelled by $\x$. We consider a non-interacting Hamiltonian $\H(A_0,\A)$ in the presence of an electric potential $A_0$ and a vector potential $\A$. Single-particle wave functions $\psi(\x)$ may have $n$ different components (spin, sublattice, $\dots$) per site.
Time-reversal can be represented by an anti-unitary operator $\T=\hat U K$, where the unitary operator $\hat U$ acts locally on the $n$ internal degrees of freedom, and $K$ denotes complex conjugation. 
TRS of a  Hamiltonian is defined by $\T\H(A_0,\A)\T^{-1}=\H(A_0,-\A)$, i.e.\  reversing the direction of electronic motion has the same effect as reversing the magnetic field.

We choose a chemical potential $\mu=0$, such that at zero temperature, the expectation value of the charge density is given by
\begin{equation}
    \label{eq-charge-density-app}
    \<\rr(\x)\>_{A_0,\A} = (-e)\sum_{E_\alpha \leq 0} |\eta_\alpha(\x)|^2
,\end{equation}
where $E_\alpha$ are the energy eigenvalues of  $\H(A_0,\A)$, and $\eta_\alpha(\x)$  the corresponding eigenfunctions. Due to the locality property of $\T$ as described above, the probability  at every site is unchanged under a time-reversal transformation, that is $|\T\eta_\alpha(\x)|^2=|\eta_\alpha(\x)|^2$. 
For a TRS Hamiltonian, $\T\eta_\alpha(\x)$ is an eigenstate of $\H(A_0,-\A)$ with energy $E_\alpha$, and applying time-reversal to Eq.~\eqref{eq-charge-density-app}, we find that the charge density is independent of the sign of the magnetic field,
\begin{equation}
    \label{eq-density-time-reversal}
    \<\rr(\x)\>_{A_0,\A} = \<\rr(\x)\>_{A_0,-\A} \ .
\end{equation}
In other words, the charge response is TRS.

\subsection{Low-energy states of a bulk TI in the presence of a magnetic field}
Usually, the low-energy states of a TI are assumed to be localized on the surface. However, in the literature, a wormhole effect has been discussed~\cite{Rosenberg:2010dj}, whereby a flux tube threaded through a TI can lead to zero-energy states in the bulk.
However, this effect requires that the diameter of the flux tube is much smaller than the lattice spacing, and we now argue that it does not occur when the diameter is larger and a continuum approximation applies.

For simplicity, we consider a geometry of a partially infinite TI that is bounded in $z$-direction by two coordinate planes. Then, the representative continuum bulk Hamiltonian can be written as a sum of two anticommuting self-adjoint operators, $\H = \hat X + \hat Y$, $\{\hat X,\hat Y\}=0$, where $\hat X = \tau_z(\sigma_x(p_x - e\A_x)+\sigma_y(p_y - e\A_y))$ and $\hat Y=\tau_z\sigma_z p_z + \tau_x m(z)$. Here, $\A=(\A_x,\A_y,0)$ is the vector potential modeling a flux tube along the $z$-axis, $p_\mu$ is the momentum operator, $\sigma_\mu$, $\tau_\mu$ are Pauli matrices acting on spin and valley degrees of freedom and $m(z)$ is the mass profile that interpolates between $-M$ inside and $+M$ outside the material. It follows that $\hat H^2 = \hat X^2 + \hat Y^2$ and that $\hat X^2$ and $\hat Y^2$ commute. Thus, any low-energy state of $\H^2$ must be a low-energy state of $\hat Y^2$, but the latter describes a TI in 1D whose low-energy states are localized on the boundary in $z$-direction. Hence, there are no low-energy bulk states for the 3D Hamiltonian.
We expect that this result is generic and also applies to a spherical geometry. It certainly agrees with a numerical calculation of the eigenstates in a cubical geometry, Fig.~\ref{figure-spectrum}(b) of the main text.

\subsection{Continuum model: Sphere with flux tube}
In this section, we detail the calculation of the eigenstates of the Hamiltonian \eqref{eq-hamiltonian} for a thin flux tube.

First, note that the Dirac Hamiltonian~\eqref{eq-hamiltonian} is written in a spin basis that rotates with the angular coordinates.
Here, the wave function acquires a phase of $\pi$ when the azimuthal coordinate makes a full rotation. Hence, the wave function has to satisfy antiperiodic boundary conditions, $\psi(\theta,\phi+2\pi)=-\psi(\theta,\phi)$, which means that the angular momentum is half-integral, $m=\pm\frac12,\pm\frac32,\dots$.
In principle, it would also be possible to use a fixed spinor basis and periodic boundary conditions, but then the Hamiltonian would be different from Eq.~\eqref{eq-hamiltonian}. 
However, the latter form of the Hamiltonian informs the boundary conditions, which are that $|\psi(\theta,\phi)|^2$ says finite as $\theta\to0,\pi$.

The Hamiltonian $\H$ is purely off-diagonal, hence its square is a diagonal matrix, $\H^2 = \text{diag}(h^+h^-,h^-h^+)$. We find
\begin{align}
    h^+h^-
    =
    &-\frac1{\sin\theta}\del_\theta(\sin \theta \del_\theta)
    + \left(\frac{i\del_\phi}{\sin \theta}
        + \frac12 \cot \theta + e R\A_\phi\right)^2
    - \frac1{\sin \theta}\del_\theta(\sin \theta\, eR\A_\phi)
    + \frac12
\end{align}
if the vector potential $\A_\phi=\A_\phi(\theta)$ does not depend on the azimuthal angle $\phi$.
This formula is a special case of the Schr\"odinger--Lichnerowicz formula~\cite{Lichnerowicz:1963vc}. The first two terms correspond to the spinor Laplacian, where the term $\frac12 \cot \theta$ captures the spinorial nature of the wave function. The third term is a Zeeman term and the constant $\frac12$ corresponds to the scalar curvature of the sphere.

We now proceed with the eigenvalue problem for the operator~\eqref{eq-hamiltonian-squared}, $h^+h^-\psia(x)=E^2\psia(x)$, where we have substituted $x=\cos\theta$.
In any region where the flux is a linear function, $N_\phi(x)=Ax+B$, the eigenvalue equation can be brought into the form of a hypergeometric differential equation by making the ansatz
$\psia(x) = (1-x)^\alpha (1+x)^\beta g(x)
$.
Choosing the exponents
\begin{align}
    \alpha = \eta \frac12\left(m -\frac12 - A - B\right)
,\quad
    \beta = \eta \frac12\left(m + \frac12 + A - B\right)
,\quad
    \eta = \pm 1
\end{align}
and using new coordinates $\xi=(1-x)/2$, we find that the eigenvalue equation is equivalent to
\begin{align}
    \Bigg[
        \xi(1-\xi)\frac{d^2}{d\xi^2}
        + \eta \Bigg(
            m-B-A+\eta-\frac12
        &- \left[2\left(m - B + \eta\frac12\right)+ \eta\right] \xi
        \Bigg) \frac{d}{d\xi}
        - \left[\left(m - B + \eta\frac12\right)^2 - A^2 - E^2 \right]
    \Bigg]g(\xi) = 0
    \label{eq-eigenvalue-hypergeometric}
.\end{align}
This has the form of a \emph{hypergeometric differential equation}~\cite{NIST:DLMF}
\begin{equation}
    \xi(1-\xi)\frac{d^2 g}{d\xi^2} + [c - (a+b+1)\xi]\frac{dg}{d\xi} - ab\,g = 0
\end{equation}
with parameters $a,b,c\in \R$. For the positive sign, $\eta=+1$, we can identify the parameter combinations
\begin{equation}
    d := m-B+1/2
    ,\quad
    c = d-A
    ,\quad
    a+b = 2d
    ,\quad
    ab = d^2 - A^2 - E^2
.\end{equation}
Since we have modeled the flux tube by a piecewise linear function $N_\phi(x)$, we can make a piecewise ansatz for the wave function. For positive angular momentum $m\geq 1/2$, we use $\eta=+1$ and consider
\begin{equation}
    \label{eq-wave-function-ansatz}
    \psia(x) = \begin{cases}
        (1-x)^{\frac12(m-\frac12)} (1+x)^{\frac12(m+\frac12) - \frac{N_0}{\delta}} g_0(x)
            &\text{ if } 1-\delta \leq x \leq 1 \\
        (1-x)^{\frac12(m - N_0-\frac12)} (1+x)^{\frac12(m - N_0+\frac12)} g(x)
            &\text{ if } -1+\delta \leq x \leq 1-\delta \\
        (1-x)^{\frac12(m-\frac12) - \frac{N_0}{\delta}} (1+x)^{\frac12(m+\frac12)} g_1(x)
            &\text{ if } -1 \leq x \leq -1+\delta
    .\end{cases}
\end{equation}
The eigenvalue equation for the wave function turns into one hypergeometric equation for each of the functions $g,g_0,g_1$. The choice $\eta=+1$ ensures that the wave function $\psia(x)$ is finite at the poles if and only if the functions $g_0(x)$ and $g_1(x)$ are finite at $x=+1$ and $x=-1$, respectively.

The continuity of the wave function $\psia$ and its derivative $\del_x \psia$ lead to jump conditions for the functions $g_0,g,g_1$ and their derivatives at the points $x=\pm(1-\delta)$. We change coordinates to $\xi=(1-x)/2$, so that we consider functions $g(\xi),g_0(\xi),g_1(\xi)$ with derivative $g'(\xi) = \del_\xi g(\xi) = -2\del_x g(x)|_{\xi=(1-x)/2}$. Setting $\eps=\delta/2$, we find that
\begin{equation}
    \label{eq-jump-full}
    \frac{g'(\eps)}{g(\eps)} - \frac{g'_0(\eps)}{g_0(\eps)}
    = \frac{N_0}{\eps}
    ,\text{ and }\quad
    \frac{g'_1(1-\eps)}{g_1(1-\eps)} - \frac{g'(1-\eps)}{g(1-\eps)}
    = \frac{N_0}{\eps}
.\end{equation}

We can now solve Eq.~\eqref{eq-eigenvalue-hypergeometric} using hypergeometric functions. Let $F(a,b,c;\xi) \equiv {}_2F_1(a,b,c;\xi)$ denote the Gauss hypergeometric function.
We introduce the abbreviation $k=m+\frac12=1,2,3,\dots$ and let $c=k-N_0$. For simplicity, we consider only the case where the latter quantity is \emph{not} an integer, $c\not\in\Z$. Then, the general solution is
\begin{align}
    \label{eq-g0}
    g_0(\xi) &=
        \alpha_0 F
        \left( k-\frac{N_0}{2\eps} + \sqrt{\left(\frac{N_0}{2\eps}\right)^2 + E^2}
        , k-\frac{N_0}{2\eps} - \sqrt{\left(\frac{N_0}{2\eps}\right)^2 + E^2}
        , k
        ; \xi\right)
\\
    g(\xi) &= \alpha w_1\left(\xi\right) + \beta w_2\left( \xi \right)
    \nonumber
\\* &=
    \label{eq-solution-middle}
        \alpha F\left(c - E, c + E, c, \xi\right)
    +
        \beta \xi^{1-c} F\left(1 - E, 1+E, 2-c, \xi\right)
\\
    \label{eq-g1}
    g_1(\xi) &=
        \alpha_1 F
        \left(k - \frac{N_0}{2\eps} + \sqrt{\left(\frac{N_0}{2\eps}\right)^2 + E^2}
        , k - \frac{N_0}{2\eps} - \sqrt{\left(\frac{N_0}{2\eps}\right)^2 + E^2}
        , k+1
        ; 1-\xi\right)
\end{align}
where $\alpha_0,\alpha,\beta,\beta_0 \in \C$ are complex parameters to be determined, and where we have already used that the solutions $g_0(\xi)$ and $g_1(\xi)$ have to be finite at the points $\xi = 0$ and $\xi=1$, respectively. Here, $w_1(\xi)$ and $w_2(\xi)$ are two linearly independent solutions of the hypergeometric differential equation. We can also consider two other, linearly independent solutions $w_3(\xi)$ and $w_4(\xi)$, and write the middle part as
\begin{align*}
    \numberthis
    g(\xi)
    &= \tilde{\alpha}w_3(\xi) + \tilde{\beta}w_4(\xi)
=
    \tilde{\alpha} F\big(c-E,c+E, c+1; 1-\xi\big)
    +\tilde{\beta} (1-\xi)^{-c} F\big( 1-E,\ 1+E,\ 1-c; 1-\xi \big)
\end{align*}
where the parameters $\tilde{\alpha},\tilde{\beta}$ are related to $\alpha,\beta$ by \emph{Kummer's relations}~\cite{NIST:DLMF}
\begin{align}
    \tilde{\alpha}
    =
    \alpha \frac{\Gamma(c)\Gamma(-c)}{\Gamma(E)\Gamma(-E)}
    + \beta \frac{\Gamma(2-c)\Gamma(-c)}{\Gamma(1-c + E)\Gamma(1-c - E)}
,\quad
    \tilde{\beta}
    =
    \alpha \frac{\Gamma(c)\Gamma(c)}{\Gamma(c-E)\Gamma(c+E)}
    + \beta \frac{\Gamma(2-c)\Gamma(c)}{\Gamma(1+E)\Gamma(1-E)}
.\end{align}
Here, $\Gamma(z)$ is the Gamma function.

The parameters are determined by the jump conditions. Expanding to leading order in $\eps$, we obtain
\begin{align}
    0 &= \alpha\left[\frac{c^2-E^2}{c}
        - \left(\frac{g_0'(\eps)}{g_0(\eps)} + \frac{N_0}\eps\right)
        + O(\eps) \right]
    + \beta \left[(1-c)\eps^{-c}
        - \left(\frac{g_0'(\eps)}{g_0(\eps)} + \frac{N_0}\eps\right) \eps^{1-c}
        + O(\eps^{1-c}) \right]
\\
    0 &= \tilde{\alpha}\left[
        -\frac{c^2-E^2}{c+1}
        - \left(\frac{g_1'(1-\eps)}{g_1(1-\eps)} - \frac{N_0}\eps\right)
        + O(\eps) \right]
    + \tilde{\beta} \left[
        c\eps^{-c-1}
        - \left(\frac{g_1'(1-\eps)}{g_1(1-\eps)} - \frac{N_0}\eps\right) \eps^{-c}
        + O(\eps^{-c}) \right]
.\end{align}
To make further progress, we need to know the asymptotic expansions of the quotients $g_0'(\eps)/g_0(\eps)$ and $g_1'(1-\eps)/g_1(1-\eps)$ in the limit $\eps\to 0$.
These will be derived in the next section. Using Eqs.~\eqref{eq-asymptotics-g0} and \eqref{eq-asymptotics-g1}, we find that to leading order in $\eps$, the energy $E$ is an eigenvalue if and only if the system of linear equations
\begin{align}
    0 =
    \alpha \left[ -r_0 E^2 \right] + \beta (1-c)\eps^{-c}
,\quad
    0 =
    \tilde{\alpha}
    + \tilde{\beta} \left[-\frac{c}{p_1 - N_0}+1\right] \eps^{-c}
\end{align}
has a non-trivial solution, where $r_0$ and $p_1$ are coefficients featuring in the asymptotic expansion.
We can express $\tilde{\alpha},\tilde{\beta}$ in terms of $\alpha,\beta$. Then, the existence of a solution corresponds to the vanishing of a determinant to leading order in $\eps$. Depending on the sign of $c$, there are two possibilities for what the leading order in $\eps$ is.

\emph{Case $c>0$.} We have $\eps^{-c} \gg \eps^0$. To leading order, the determinant vanishes if
\begin{equation}
    0 = \frac{\Gamma(c)\Gamma(c)}{\Gamma(c-E)\Gamma(c+E)} · \left[-\frac{c}{p_1 - N_0} + 1\right]
.\end{equation}
According to Eq.~\eqref{eq-asymptotics-g1}, the second factor is non-zero for $c>0$, so this can only happen if the product of Gamma functions in the denominator has a pole, that is only if
\begin{align}
    E = \pm(c + n)
    ,\quad\text{ with } n = 0,1,2,\dots
    \text{ for } c = m + 1/2 - N_0 > 0
.\end{align}
In this case, the first equation implies that the second parameter is of order $\beta \sim \alpha\eps^c$. Hence, the actual solution \eqref{eq-solution-middle} is dominated by the first term,
\begin{align*}
    \label{eq-solution-c-bigger-0}
    g(\xi) &\sim  F(c-E, c+E, c; \xi) + O(\eps^{c})
\numberthis
    = F(-n, 2c+n, c; \xi) + O(\eps^{c})
    \quad\text{ for } c>0, E = c+n
.\end{align*}
This is actually a polynomial in the variable $\xi$.

\emph{Case $c<0$.} We have $\eps^0 \gg \eps^{-c}$. To leading order, the determinant vanishes if
\begin{equation}
    0 = \left(-r_0 E^2 \right) · \frac{\Gamma(2-c)\Gamma(-c)}{\Gamma(1-c+E)\Gamma(1-c-E)}
.\end{equation}
Since Eq.~\eqref{eq-asymptotics-g0} establishes that $r_0\neq 0$ whenever $c<0$, the solutions to this condition are
\begin{align}
    \label{eq-spectrum-with-zero}
    E = 0 \quad\text{ or }\quad
    E = \pm(1 - c + n)
    ,\quad\text{ with } n = 0,1,2,\dots
    \text{ for } c = m + 1/2 - N_0 < 0
.\end{align}
In other words, there is a zero-energy mode and a set of equally spaced energy modes. For the latter, we see that the second coefficient $\beta \sim \alpha\eps^{c}$ dominates the first, so the solution has the form
\begin{align}
    g(\xi)
    &\sim
    \xi^{1-c} F(c-n, 2+n - c, 2-c; \xi) + O(\eps^{|c|})
    \quad\text{ for } c<0, E = 1-c+n
    \nonumber
.\end{align}
On the other hand, for the zero-energy mode $E=0$, we see that to leading order, the first equation implies that the parameter $\beta$ has to vanish, while the parameter $\alpha$ can be any constant. If this is the case, then the second equation is satisfied to any order lower than $\eps^{-c}$. Hence, the zero-energy mode of the operator $h^+h^-$ has the shape
\begin{align}
    g(\xi)
    &\sim
    F(c, c, c; \xi) + O(\eps^{|c|})
=
    (1-\xi)^{-c} + O(\eps^{|c|})
    \quad\text{ for } c<0, E = 0
.\end{align}

\subsection{Asymptotics of hypergeometric functions}
\label{appendix-asymptotics}

In this section, we want to establish asymptotic expansions for the quotients $g'_0(\eps)/g_0(\eps)$ and $g'_1(1-\eps)/g_1(1-\eps)$, Eqs. \eqref{eq-g0} and \eqref{eq-g1}, in the limit $\eps\to 0$. These quotients feature in the jump conditions for the eigenfunctions.

The derivative of a hypergeometric function is given by
$\frac{d}{d\xi} F(a,b,c;\xi) = \frac{ab}c F(a+1,b+1,c+1;\xi)$~\cite{NIST:DLMF}.
Setting
\begin{equation}
    a = k - \frac{N_0}{2\eps} + \sqrt{\left(\frac{N_0}{2\eps}\right)^2 + E^2}
    ,\quad
    b = k - \frac{N_0}{2\eps} - \sqrt{\left(\frac{N_0}{2\eps}\right)^2 + E^2}
,\end{equation}
we find that
\begin{align}
    \label{eq-expand-g0}
    \frac{g'_0(\eps)}{g_0(\eps)}
    &=
    + \left(\frac{k^2 - E^2}{k} - \frac{N_0}{\eps} \right)
     \frac{F(a+1,b+1,k+1;\eps)}{F(a,b,k;\eps)}
\\
    \label{eq-expand-g1}
    \frac{g'_1(1-\eps)}{g_1(1-\eps)}
    &=
    - \frac{k^2 - k\frac{N_0}{\eps}-E^2}{k+1} \frac{F(a+1,b+1,k+2;\eps)}{F(a,b,k+1;\eps)}
.\end{align}
In the limit $\eps \to 0$, there are two cases: If $N_0>0$, then we have $a = k + O(\eps)$ and $b = -\frac{N_0}\eps + k + O(\eps)$, whereas if $N_0<0$, these two equations have to be interchanged.
In either case, one of the parameters $a,b$ diverges, while the argument $\xi$ of the hypergeometric functions tends to zero. This limit can be expressed in terms of \emph{Kummer's confluent hypergeometric function} $M(b,c;\xi)$, which satisfies
$
    \lim_{b\to \infty} F(a,b,c;\xi/b) = M(a,c;\xi) 
$.~\cite{NIST:DLMF}
Note that this limit is finite, so the leading term in the asymptotic expansions of the quotients $g_0'/g_0$ and $g_1'/g_1$ must be of order $\eps^{-1}$. In general, we expect that the asymptotic expansion has the form $\frac{p}\eps + q + o(1)$ where the expansion coefficients $p,q$ may still depend on the values of $k,N_0$ and $E$.

First, let us show that the term of order $\eps^{-1}$ in the expansion of $g'_0/g_0$ at the north pole is precisely $N_0/\eps$.
If $N_0>0$, the parameter $b$ diverges as $-N_0/\eps$. We write $F(a,b,k;\eps) \sim F(a,b,k;-N_0/b)$ and apply the confluent hypergeometric function, finding
\begin{equation}
    \frac{g'_0(\eps)}{g_0(\eps)}
    = -\frac{N_0}{\eps}\frac{F(a+1,b+1,k+1;\eps)}{F(a,b,k;\eps)} + o(\eps^{-1})
    = -N_0\frac{M(k+1,k+1;-N_0)}{M(k,k;-N_0)}\frac1{\eps}
    + o(\eps^{-1})
    = -\frac{N_0}{\eps} + o(\eps^{-1})
.\end{equation}
The last equality follows from the identity $M(n,n;z) = e^{z}$ for any real $n$. Since the hypergeometric function is symmetric in the parameters $a$,$b$, the same argument applies to the case $N_0<0$. This establishes the leading order.

For the expansion at the south pole, a similar argument shows that
\begin{equation}
    \frac{g'_1(1-\eps)}{g_1(1-\eps)}
    = \frac{k}{k+1}N_0\frac{M(k+1,k+2;-N_0)}{M(k,k+1;-N_0)}\frac1{\eps}
    + o(\eps^{-1})
.\end{equation}
In this case, the confluent hypergeometric function can no longer be expressed in terms of elementary functions. We are not so much interested in the precise form of the expansion, but only that the coefficient is distinct from $N_0$. We want to establish that it has the form
\begin{equation}
    \label{eq-asymptotics-g1}
    \frac{g'_1(1-\eps)}{g_1(1-\eps)}
    = \frac{p_1}{\eps} + o(\eps^{-1})
    ,\text{ with }
    p_1\equiv p_1(N_0,k) \neq N_0
    \text{ if } N_0\neq 0
    ,\text{ and }
    p_1 - N_0 < c
    \text{ if } c>0
.\end{equation}
For this, we resort to a numerical calculation: The following two plots of $p_1(N_0,k)$ for different values of $k=1,2,\dots$ confirm that the coefficient satisfies $p_1\neq N_0$ as long as $N_0\neq 0$ and that it also satisfies $p_1-N_0 < c$ if $c>0$.
\begin{center}
    \includegraphics[width=0.4\textwidth]{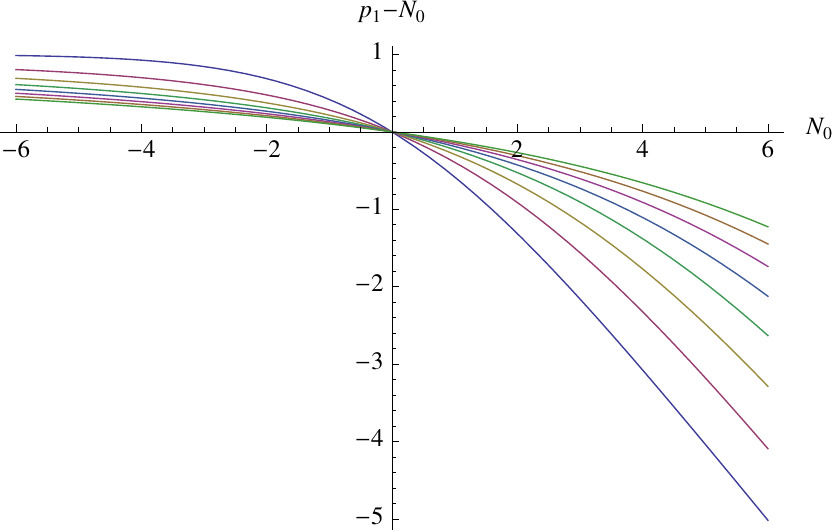}%
    \includegraphics[width=0.4\textwidth]{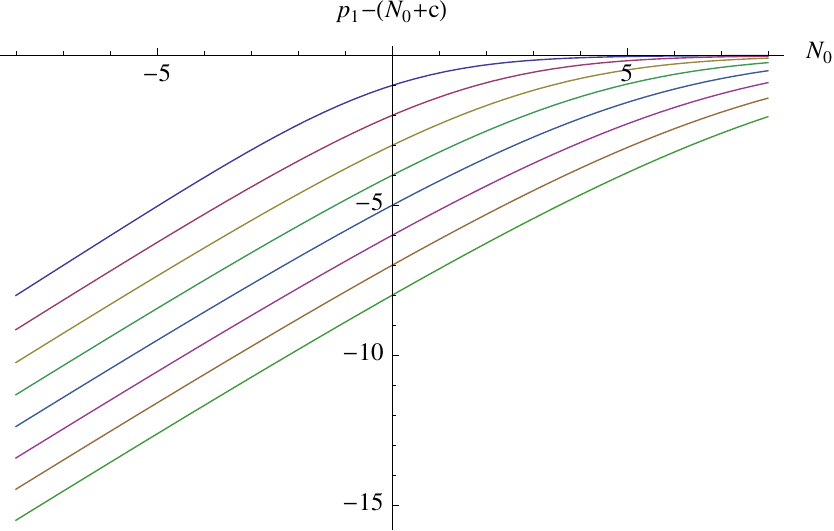}
\end{center}

At the north pole, we also need the subleading term. In this case, the limit of the hypergeometric functions is more difficult to compute, and we resort to a numerical computation again. It turns out that the asymptotic expansion depends on the sign of $c=k-N_0$, i.e.\ whether the flux exceeds the angular momentum or not. We have
\begin{equation}
    \label{eq-asymptotics-g0}
    \frac{g'_0(\eps)}{g_0(\eps)}
    = -\frac{N_0}{\eps} + \frac{c^2 - E^2}{c} + r_0 E^2+ o(1)
    ,\text{ with }
    r_0 \equiv r_0(k,N_0,E) \neq 0 \text{ if } c < 0
.\end{equation}
This is confirmed by the following plot, which shows the remainder $r_0(k,N_0,E)$ for several angular momenta $k=1,2,\dots$:
\begin{center}
    \includegraphics[width=0.7\textwidth]{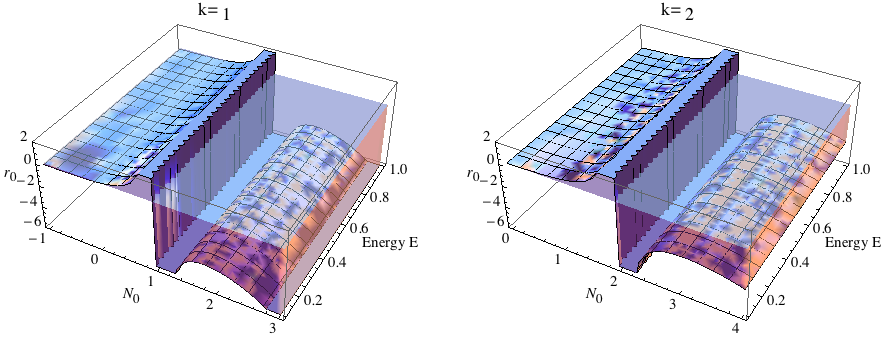}
    \includegraphics[width=0.7\textwidth]{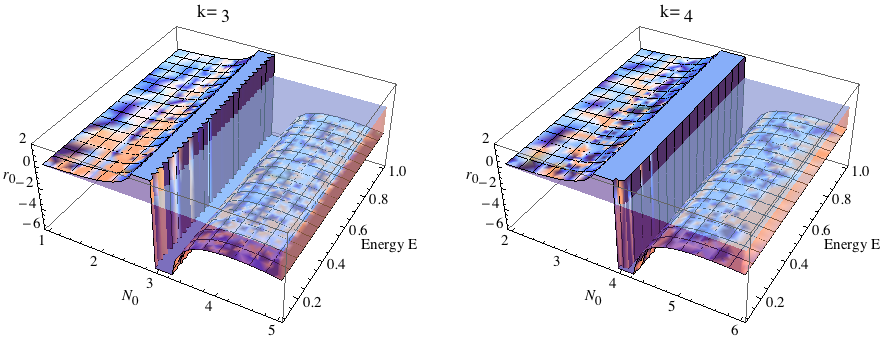}
\end{center}
Whenever $c<0$, this value is clearly negative, while for $c>0$, it can approach zero. The plot was calculated for a fixed value of $\eps=10^{-4.5}$, which is a good representation of the limit $\eps\to 0$ if the other quantities are significantly larger. Due to numerical difficulties, the value of $r_0$ at $E=0$ has to be extrapolated from its values at finite energies, but this can be done visually, as $r_0(k,N_0,E)$ is a constant function of $E$ for $c<0$.
 
The existence of the zero-energy state, Eq.~\eqref{eq-spectrum-with-zero}, crucially depends on the fact that the asymptotic expansion contains the non-vanishing term $r_0E^2$ whenever the flux exceeds the angular momentum, $c<0$.

\subsection{Lattice model: Coupling to electromagnetic gauge field}
We now couple the lattice Hamiltonian~\eqref{eq-hamiltonian-lattice} from the main text to the electromagnetic gauge field $(A_0,\A)$ in order to numerically calculate its charge response.

Peierls substitution provides a minimal gauge-invariant coupling. With link variables $\A_{\x,\x+ae_\mu} = \int_{\x}^{\x+ae_\mu} \A(\x')\cdot d\x'$, we can introduce covariant lattice derivatives
\begin{align}
    (D_\mu \psi)(\x) &= \frac1{2a} \Big[
        e^{-ie \A_{\x,\x+ae_\mu}}\psi(\x + ae_\mu)
        - e^{-ie \A_{\x,\x-ae_\mu}}\psi(\x-a e_\mu)
        \Big]
\\
    (\Laplace \psi)(\x) &= \sum_{\mu=1}^3\frac1{a^2}
        \Big[
        e^{- ie \A_{\x,\x+ae_\mu}}\psi(\x + ae_\mu)
        - 2\psi(\x)
        - e^{-ie \A_{\x,\x-ae_\mu}}\psi(\x - ae_\mu)
        \Big]
,\end{align}
where $e_\mu$ denotes the unit vector in direction $\mu=1,2,3$ and $a$ is the lattice spacing, which we set to $a=1$. Our single-particle Hamiltonian $\H(A_0,\A)$ is then given by
\begin{align}
    (\H(A_0,\A)\psi)(\x) &= -2\lambda \sum_{\mu=1}^3 \tau_z\sigma_\mu (-iD_\mu\psi)(\x)
    + \tau_x \left[(\kappa-6t)\psi(\x) - t(\Laplace\psi)(\x)
    - t\sum_{\mu=1}^3 e\mathbf{\tilde{B}}_\mu(\x)\sigma_\mu \psi(\x)
    \right]
    + eA_0(\x)\psi(\x)
.\end{align}
Here, $\lambda, \kappa, t$ are real parameters and $a^2\mathbf{\tilde{B}}_\mu(\x)$ is the magnetic flux through a square-shaped surface of side length $a$ centered at the lattice position $\x$ and perpendicular to the direction $e_\mu$. We imagine that the four components of the fermion wave function are a product of two ``orbital'' (sublattice) and two ``spin'' degrees of freedom. Then, the Pauli matrices $\sigma_\mu$ and $\tau_\mu$ act only on the spin, resp.\ orbital degrees of freedom. In the parameter range $2t \leq \kappa \leq 6t$, this model describes a (strong) topological insulator \cite{Qi:2008eu}.

Note, however, that the coupling to the electromagnetic gauge field is \emph{not} unique: Different terms can be added as long as they vanish when $\A=0$; they correspond to different $g$-factors for the electron. Here, we have chosen a contribution involving the field strength $\mathbf{\tilde{B}}_\mu(\x)$ such that the mass term (the term multiplying $\tau_x$) commutes with the square of the kinetic term, i.e.~with $(\sum_{\mu=1}^3 \sigma_\mu D_\mu)^2$, in the continuum limit ($a\to 0$).

\end{document}